High-field and low field magnetoresistance of CoFe nanoparticles

elaborated by organometallic chemistry

Reasmey P. Tan, Julian Carrey, Marc Respaud,

LPCNO-IRSAMC-INSA-UPS-CNRS, 135, av. de Rangueil, 31077 Toulouse cedex 4, France

Céline Desvaux, Philippe Renaud

Freescale Semiconductor, le Mirail B.P. 1029, 31023 Toulouse cedex, France

Bruno Chaudret

Laboratoire de Chimie de Coordination-CNRS, 205 route de Narbonne, 31077 Toulouse cedex 4,

France

**Abstract :**

We report on magnetotransport measurements on CoFe nanoparticles surrounded by an insulating organic layer. Samples were obtained by evaporating a solution of nanoparticles on a patterned substrate. Typical behaviour of Coulomb blockade in array of nanoparticles is observed. High and low field magnetoresistance have been evidenced. Below 10 K, a large high-field magnetoresistance is measured, reaching up to 500 %. Its amplitude decreases strongly with increasing voltage. At 1.6 K, this high-field magnetoresistance vanishes and an inverse low field tunnelling magnetoresistance is observed.

**Main text :**

Extensive studies have been devoted to nanoparticles (NPs) arrays, especially when NPs are magnetic, both for the fundamental aspects and for the potential applications based on their magnetoresistive properties.[1] Large scale production of ferromagnetic NPs surrounded by an organic layer is nowadays possible by different chemical routes, but their integration and their magnetotransport properties are not well controlled. In such hybrid systems containing an assembly of ferromagnetic metallic NPs surrounded by an insulating organic barrier, two types of magnetoresistance (MR) are listed. The first one, the tunnel MR, is linked to the relative orientation of the magnetization of spin-polarized neighbouring NPs.[2] Experimental observations of this MR have been reported.[3-8] An estimate of the spin polarization of the NPs can be deduced from the amplitude of the tunnel MR.[4-6, 8] The second one is a high-field MR effect which occurs above the saturation field of the ferromagnetic NPs.[9-14] In some cases, huge values of MR are measured [15, 16] (MR > 1000 %) and hypotheses such as the contribution of magnetic impurities dispersed in the insulating barrier [9] or of spin disorder at the surface of the particles [5, 12, 17, 18] have been proposed, although the origin of this high-field behaviour remains unclear.

In previous works,[16, 19] we have reported on magnetotransport properties of millimetre long three-dimensional super-lattices of CoFe nanoparticles surrounded by organic ligands. As-prepared super-lattices exhibit various magnetotransport properties and especially three different types of MR. The first one, between 1.8 K and 10 K, is a high-field MR with huge amplitude, up to 3000 % in optimized temperature and bias voltage conditions. Besides, the MR amplitude shows a strong dependency on the applied voltage and temperature. Below 1.8 K, the high-field MR vanishes and tunnel MR develops. The transition between both regimes of MR has been found to be abrupt.[16, 19]. Finally, an original mechanism of MR related to the Coulomb blockade properties of the samples has also been found.[19] These as-prepared super-lattices are millimetric and cannot be integrated directly into electronic devices. Therefore, in this article, we report on the preparation and the study of the magnetotransport properties of CoFe NPs deposited onto patterned substrates. We show that the behaviour of the nanodevices is very similar to those of as-prepared super-lattices, except a typical 2D character of the Coulomb blockade properties instead of a 3D one. In particular, the first two types of MR mentioned above are observed. The high-field MR displays a strong dependency on the applied voltage and its amplitude reaches up to 500 %.

The devices we studied were prepared as follows. We first synthesized CoFe NPs according to our usual procedure. Solid super-lattices contain a regular network of CoFe nanoparticles displaying a diameter of 15 nm and surrounded by a mixture of oleic acid and hexadecylamine (for the synthesis method and detailed characterization, see Ref. 20). The super-lattices are then dissolved in THF, in order to get a colloidal solution of NPs. A droplet of this solution is deposited onto a pre-patterned substrate of Si/SiO$_2$ layers in a glove box, under Ar atmosphere. This procedure guarantees a complete dissolution of the 3D supercrystal and thus keep the same narrow size distribution on the substrate than the one originally present in the super-lattices. Each substrate carries 8 pairs of interdigitated (finger-like shape) Au electrodes made by optical lithography. The width between two fingers of the interdigitated electrodes is 5 µm. Considering the NPs size (15 nm in diameter), approximately 300 NPs are placed in series between the electrodes. Once the solvent is evaporated, the time to transfer the sample from the glove box to the inside of the cryostat was kept to a minimum (few tens of seconds) to avoid any oxidation. Magnetotransport measurements were performed in a Cryogenic cryostat equipped with a superconducting coil (up to 12 T) using a Keithley 6430 sub-femtoamp sourcemeter. Over the 8 interdigitated electrodes of the sample, 3 of them do not show any measurable resistance. The electrodes number 2, 3, 5, 6 and 8 exhibit resistance between 25 kΩ and 430 kΩ at room temperature. This dispersion in resistance may come from the non-symmetric position of the droplet and from its inhomogeneous evaporation during the sample elaboration. As depicted in the inset of Fig.1a, one or two monolayers of disordered NPs are obtained.

Current-voltage [$I(V)$] characteristics of the different electrodes present typical features of Coulomb blockade in an array of NPs. $I(V)$ characteristics measured at room temperature exhibit ohmic behaviour (see Fig. 1a). When the temperature is lowered, the $I(V)$ curves becomes non linear and a Coulomb gap appears (see Fig. 1b). In NPs assemblies, $I(V)$ characteristics in the Coulomb blockade regime are expected to follow the relation $I \propto [(V - V_T)/V_T]^\zeta$, where $V_T$ is the threshold voltage above which conduction occurs. $\zeta$ is an exponent related to the dimensionality: $\zeta = 1$ and $\zeta = 5/3$ have been calculated for 1D and 2D arrays of disordered NPs respectively.[21, 22] Fitting the $I(V)$ characteristics at 1.6 K (see Fig. 1c) allows us to extract both the value of $\zeta$ and $V_T$ (see Fig. 1d). For all the measurable electrodes, $\zeta$ ranges between 1.87 (± 0.01) < $\zeta$ < 2.12 (± 0.02).

We now turn to the high-field MR properties. In the following, the MR is defined as ($R_h$ - $R_l$) / $R_l$, where $R_h$ ($R_l$) is the high (low) resistance of the $R(H)$ characteristic. All the $R(H)$ curves come from the electrode number 5. Fig. 2a shows a typical curve of high-field MR at $T$ = 2.9 K, with a bias voltage $V$ = 1.5 V and a maximal applied magnetic field $\mu_0H$ = 2.6 T. When $V$ is lowered to 0.1 V, the MR amplitude increases significantly from 15 % ($V$ = 1.5 V) to 50 % ($V$ = 0.1 V) (see Fig. 2b). The MR($V$) dependency (see Fig. 2d) is deduced from $R(H)$ measurements and indicates a strong influence of the bias voltage. For instance, when the bias voltage is 0.06 V and the magnetic field swept up to 8.8 T a huge amplitude reaching up 500 % is measured (see Fig. 2c).

Below $T$ = 1.6 K, this high field contribution abruptly disappears. A vanishing high-field MR is observed with an amplitude which is less than 1 % / T at 2 V (see Fig. 3). In addition, a low-field inverse tunnel MR, with small amplitude (~ 1.5 %), is observed. The curve exhibits a minimum resistance at a field around 0.1 T and a saturation field around 1 T.

We now discuss the various magnetotransport properties. First of all, the electronic conduction is well described by Coulomb blockade. The values of $\zeta$ ranging from 1.87 to 2.12 are well below the one measured on 3D super-lattices ($\zeta$ = 3.75).[16] This confirms the two-dimensional character of the electronic conduction in the NP array. Indeed, even if the values of the exponent are slightly higher than the expectable theoretical values in 2D arrays, they are similar to the ones - close to 2 - reported in other experimental works.[23-25] Besides, $V_T$ values (~ 0.5 V for ~ 300 NPs measured) are coherent with our previous measurements on as-prepared super-lattices (~ 25 V for ~ 30000 NPs measured), since $V_T$ is expected to increases with the number of particles.[21, 22]

With decreasing temperature, a huge high-field MR which is very sensitive to the applied bias voltage develops below 10 K. It cannot be attributed to spin-dependent tunneling between the NPs. Indeed, even when assuming a full spin polarization of the NPs, the maximum tunnel magnetoresistance would be of only 100 %.[2] Identical high-field behaviour has been previously observed in as-prepared CoFe super-lattices, which were more extensively studied.[16,19] In this case, a universal $H/T$ dependence of the high-field MR was pointed out. We propose that this MR is due to a mechanism of electronic transfer from one particle to another through a paramagnetic localized state, localized either at the surface or in between the particles (for a more detailed discussion, see Ref. 16). Preliminary simulations based on a modified approach of the model developed by Huang et *al.*,[18] including the bias voltage influence, permit to well reproduce the field, the temperature and the bias voltage dependences of the high field MR.[26]

At low temperature, as the high field MR vanishes, a low-field MR contribution is observed. The sharp transition between both regimes of MR has been evidenced more accurately in the case of as-prepared supercrystals[19]. Besides, the *R(H)* features in this range of temperature are identical in both cases: the field corresponding to the peak of the minimum value of the resistance and the one required for the saturation are the same, but with some differences in the MR amplitude.

In conclusion, we have investigated the magnetotransport properties of ferromagnetic CoFe nanoparticles deposited onto a pre-patterned substrate. The electronic transport properties are characteristics of a 2D disordered arrays of nanoparticles in the Coulomb blockade regime. MR measurements display similar features than those performed on wider self-organized systems. This suggests that these CoFe NPs can be integrated in nanosized devices to obtain large value of MR without the need of large spin polarization. We propose that the large high field MR amplitudes are connected to the presence of some paramagnetic impurities into the organic layer. This latter point deserves much more work in order to get a definitive conclusion.

**References and notes :**

[1] G. Reiss and A. Hütten, Nature Materials **4**, 725 (2005).

[2] J. Inoue, and S. Maekawa, Phys. Rev. B **53**, R11927 (1996).

[3] C. T. Black, C. B. Murray, R. L. Sandstrom, and S. Sun, Science **290**, 1131 (2000).

[4] B. Hackenbroich, H. Zare-Kolsaraki, and H. Micklitz, Appl. Phys. Lett. **81**, 514 (2002).

[5] H. Zare-Kolsaraki and H. Micklitz, Phys. Rev. B **67**, 224427 (2003).

[6] H. Zare-Kolsaraki and H. Micklitz, Phys. Rev. B **67**, 094433 (2003).

[7] K. Yakushiji, S. Mitani, K. Takahashi, S. Takahashi, S. Maekawa, H. Imamura, and H. Fujimori, Appl. Phys. Lett. **78**, 515 (2001).

[8] T. Zhu and Y. J. Wang, Phys. Rev. B **60**, 11918 (1999).

[9] O. Chayka, L. Kraus, P. Lobotka, V. Sechovsky, T. Kocourek, and M. Jelinek, J. Mag. Mag. Mat. 300, 293 (2006).

[10] B.J. Hattink, M. García del Muro, Z. Konstantinovíc, X. Batlle, A. Labarta, and M. Varela, Phys.Rev.B **73**, 45418 (2006).

[11] W. Wang, M.Yu, M. Batzill, J. He, U. Diebold, and J. Tang, Phys.Rev.B **73**, 134412 (2006).

[12] H. Zeng, C.T. Black, R.L. Sandstrom, P.M. Rice, C.B. Murray, and Shouheng Sun, Phys.Rev.B **73**, 20402 (2006).

[13] C. Park, Y. Peng, J.-G Zhu, D.E. Laughlin, and R.M. White, J. Appl. Phys. **97**, 10C303 (2005).

[14] D. D. Sarma, S. Ray, K. Tanaka, M. Kobayashi, A. Fujimori, P. Sanyal, H. R. Krishnamurthy, and C. Dasgupta, Phys. Rev. Lett. **98**, 157205 (2007).

[15] P. Chen, D. Y. Xing, Y. W. Du, J. M. Zhu, and D. Feng, Phys. Rev. Lett. **87**, 107202 (2001).

[16] R. P. Tan, J. Carrey, C. Desvaux, P. Renaud, B. Chaudret, M. Respaud, J. Magn. Magn. Mater., accepted for publication, arXiv:0710.1750v1.

[17] M. Holdenried, B. Hackenbroich, and H. Micklitz, J. Magn. Magn. Mater. **231**, 13 (2003).

[18] Z. Huang, Z. Chan, K. Peng, D. Wang, F. Zhang, W. Zhang, and Y. Du, Phys.Rev.B **69**, 94420 (2004).

[19] R. P. Tan, J. Carrey, C. Desvaux, J. Grisolia, P. Renaud, B. Chaudret, M. Respaud, Phys. Rev. Lett. **99**, 176805 (2007).

[20] C. Desvaux, C. Amiens, P. Fejes, P. Renaud, M. Respaud, P. Lecante, E. Snoeck, and B. Chaudret, Nature Materials **4**, 750 (2005).

[21] A. A. Middleton and N. S. Wingreen, Phys. Rev. Lett. **71**, 3198 (1993).

[22] D. M. Kaplan, V. A. Sverdlov, and K. K. Likharev, Phys. Rev. B **68**, 045321 (2003).

[23] C. Kurdak, A. J. Rimberg, T. R. Ho, and J. Clarke, Phys. Rev. B **57**, R6842 (1998).

[24] R. Parthasarathy, X. Lin, and H. M. Jaeger, Phys. Rev. Lett. **87**, 186807 (2001).

[25] .K. Elteto, X. Lin, and H.M. Jaeger, Phys. Rev. B **71**, 205412 (2005).

[26] R. P. Tan, J. Carrey, and M. Respaud, submitted, arXiv:0802.1007v1.

**Figure legends**

Fig. 1 : (a) *I(V)* characteristics of the measurable electrodes at room temperature, inset: SEM picture of NPs deposited on the substrate (b) at $T = 1.6$ K (c) fitting curves of *I(V)* at $T = 1.6$ K (d) plot of exponent $\zeta$ and $V_T$ for each electrode at $T = 1.6$ K.

Fig. 2 : transport measurements on electrode 5: (a) *R(H)* characteristics at $T = 2.9$ K, $V = 1.5$ V (b) *R(H)* characteristics at $T = 2.9$ K, $V = 0.1$ V (c) *R(H)* characteristics at $T = 2.9$ K, $V = 0.06$ V (d) MR variation as a function of *V*, at $T = 2.9$ K and $\mu_0 H = 2.6$ T. The values are deduced from complete *R(H)* measurements.

Fig. 3 : transport measurements on electrode 5 : *R(H)* characteristics at $T = 1.6$ K, $V = 2$ V.

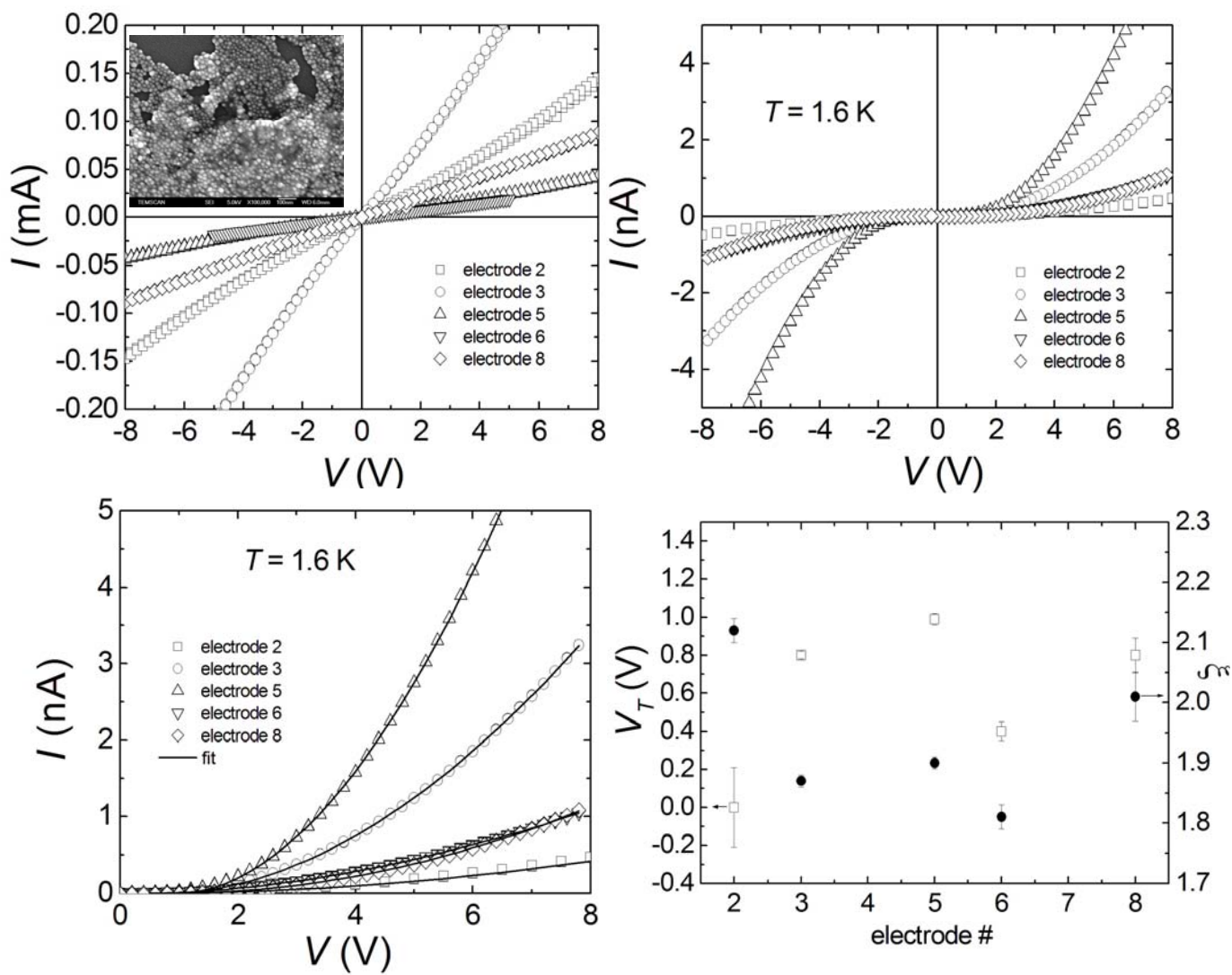

**Fig. 1**

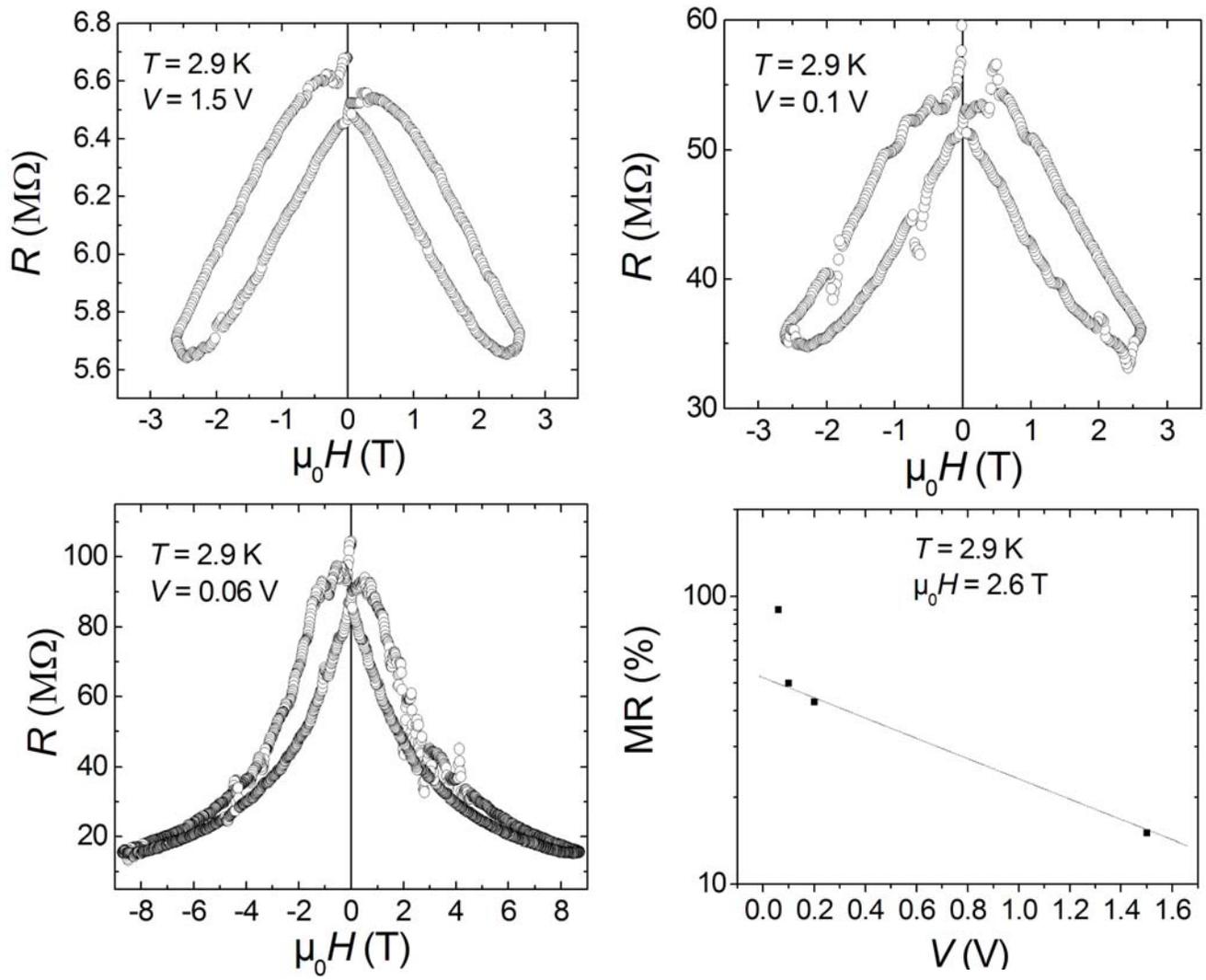

**Fig.2**

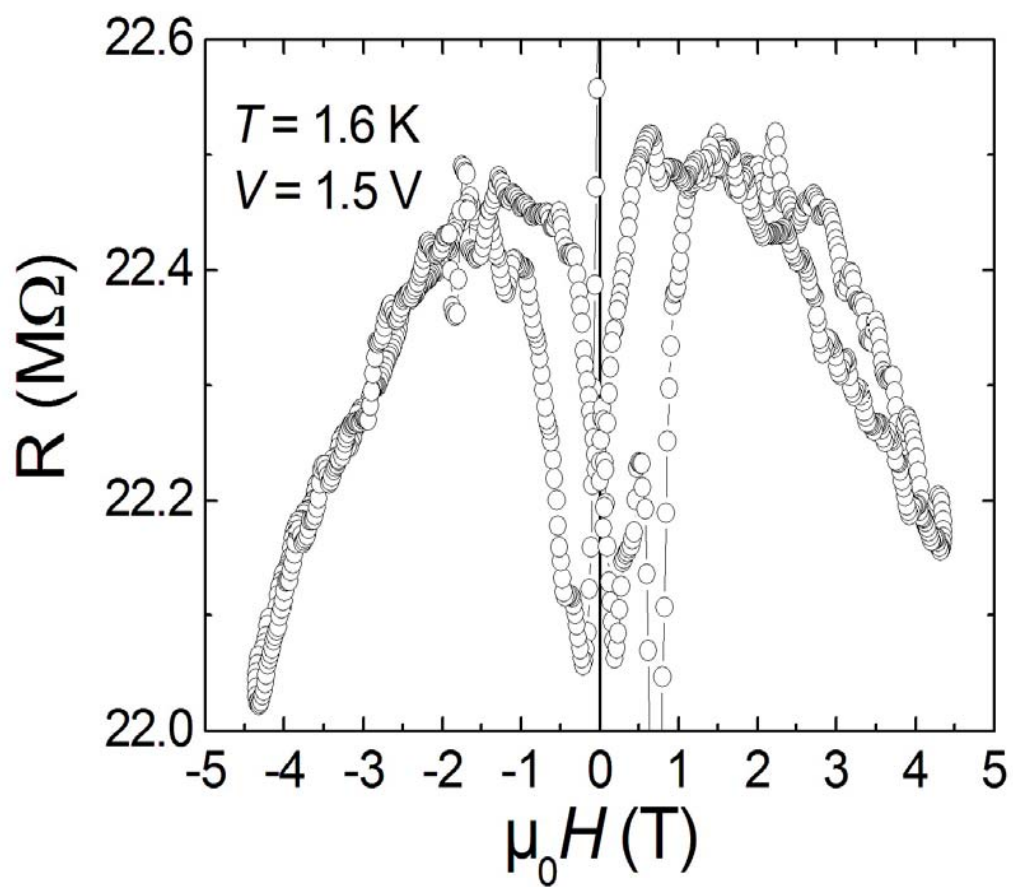

**Fig. 3**